\documentclass[prl,twocolumn,showpacs,preprintnumbers,amsmath,amssymb]{revtex4}
\usepackage{graphicx}
\usepackage{dcolumn}
\usepackage{bm}
\usepackage{textcomp}

\begin{document}

\title{Optical Spectroscopy of Ultracold Atoms on an Atom Chip}

\author{A. G\"unther}
\email{aguenth@pit.physik.uni-tuebingen.de}
\author{H. Bender}
\author{A. Stibor}
\author{J. Fort\'agh}
\author{C. Zimmermann}
\affiliation{Physikalisches Institut der Universit\"at T\"ubingen,
Auf der Morgenstelle 14, 72076 T\"ubingen, Germany}

\date{\today}

\begin{abstract}
We experimentally demonstrate optical spectroscopy of magnetically trapped atoms on an atom chip. High resolution optical spectra of individual trapped clouds are recorded within a few hundred milliseconds. Detection sensitivities close to the single atom level are obtained by photoionization of the excited atoms and subsequent ion detection with a channel electron multiplier. Temperature and decay rates of the trapped atomic cloud can be monitored in real time for several seconds with only little detection losses. The spectrometer can be used for investigations of ultracold atomic mixtures and for the development of interferometric quantum sensors on atom chips.
\end{abstract}
\maketitle
Since the early days of atomic physics, optical spectroscopy has always been
one of the most important and successful scientific tools for investigating
the structure of atoms and the properties of atomic and molecular gases.
Unfortunately, the rich arsenal of optical spectroscopy methods is difficult
to apply in the field of ultra cold atoms, in particular Bose-Einstein
condensates, atomic Fermi gases, and atomic mixtures confined in optical or
magnetic trapping potentials. This is mainly because the momentum, transferred
from the light to the atoms during spectroscopy, rapidly heats up and destroys
the atomic density and velocity distribution. Clouds of ultra cold atoms are
thus typically detected by absorption imaging after some time of ballistic
expansion. Atom numbers, density and velocity distributions at a given moment
can be derived from such images. Up to now, the overwhelming success of the
field is still almost exclusively based on this main detection tool.
Nevertheless, this method has its limitations since it still destroys the
entire sample during detection and its restricted sensitivity requires a
minimum of a few hundred atoms in the cloud. Furthermore, observation of
dynamic properties usually requires repetitive cycles of the experiment which
is time consuming and susceptible to technical
noise. Similarly, spectra can only be taken with a maximum rate of one
data point per cycle. The development of a method which allows
for direct observation of atoms inside the trap in real time
would thus be highly desirable.

This is possible with optical spectroscopy if the efficiency with which the
excited atoms can be detected is substantially enhanced, ultimately up to
single atom sensitivity. Then, only a small fraction of the cloud is spent for
detection while its main part is left undisturbed. Note that even an ideal
detector introduces unavoidable losses in the sense that it removes particles
from the clouds quantum state and projects it onto an eigenstate of the
detector. In optical spectroscopy electronically excited atoms can be detected
with unchallenged sensitivity by photo ionization and subsequent ion detection
\cite{hurst1979}. Here we apply this approach to two photon spectroscopy of
ultracold rubidium atoms on an atom chip. The trapped rubidium atoms are
excited to the $5D$ state by two photon absorption of light from a diode laser
near 778nm. An additional fiber laser near 1080nm subsequently ionizes the
excited atom \cite{kraft2007}. The ion is collected by an ion optics and
guided to a channel electron multiplier (CEM) where it is detected with
about $50\%$ probability \cite{stibor2007}. With this setup we record decay
curves and complete two photon excitation spectra with a single cloud of about
$10^{6}$ atoms. Collective properties such as temperature, density
distribution and elementary excitation can be
monitored in situ and with high resolution. The scheme is
particulary well suited to the large class of experiments where atoms are
trapped in an optical dipole trap with a laser wavelength shorter than 1250nm
such that the trapping laser simultaneously provides the light for the ionization.

\begin{figure}[tb]
\centerline{\scalebox{0.45}{\includegraphics{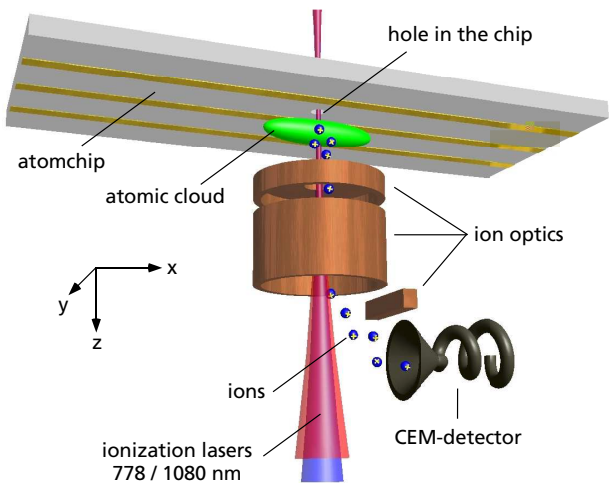}}}
\caption{(Color Online) Sketch of the experimental situation (not to scale). A cloud of rubidium atoms is placed directly underneath a hole in the atom chip. Two laser beams with wavelength of 778nm and 1080nm are focused through the bore, ionizing atoms from the ultracold atomic cloud. The generated ions are collected by an ion optics and guided to the CEM-detector.}
\label{fig:Figure1}
\end{figure}

The experimental setup is similar to our previous experiments \cite{Fortagh2003}. A cloud of typically $3\times 10^8$ $^{87}$Rb atoms is loaded from a magneto optical trap into a standard magnetic trap where the atoms are evaporatively cooled to a temperature of about 15$\mu$K. Subsequently the atoms are loaded to an atom chip which allows for magnetic trapping and positioning of the thermal cloud with sub-$\mu$m precision \cite{guenther2005a}. 
As shown in Fig.\;\ref{fig:Figure1}, the ionization takes place directly underneath a 200$\mu$m wide hole in the chip.
The magnetic trapping potential at this position is generated by the conductor geometry on the chip and characterized by the axial and radial oscillation frequency of $\omega_x=2\pi\times 17$ Hz and $\omega_r=2\pi\times 240$ Hz. The two laser beams with wavelengths of 778nm and 1080nm are aligned perpendicular to the chip surface and focussed through the hole in the chip.

Initially the atoms are prepared in the $5S_{1/2}, F=2, m_F=2$ groundstate. The multiphoton ionization process takes place in two steps \cite{kraft2007}. At first the atoms are excited into the $5D_{5/2}$-state via a resonant two photon transition driven by a grating stabilized continuous-wave diode laser with a wavelength near 778.1065nm. The transition rate is enhanced by the the intermediate $5P_{3/2}$ level which is detuned by only $2$nm. With the used maximum power of $0.5$mW and a beam radius of $34\mu$m we expect transition rates of $2250 s^{-1}$. In the second step the excited atoms are ionized by a continuous wave fiber laser at $1080$nm with a Gaussian single mode beam profile and a spectral width of about $1$nm. With a beam radius of $26\mu$m and a maximum laser power of $1.6$W the expected transition rate of 1/(70ns) is sufficiently strong to overcome the spontaneous decay rate of 1/(240ns) to the ground state via the $6P$ excited state.

Mainly due to the off resonant coupling of the $5S$ and the $5P$ states, both lasers give raise to an AC-Stark shift of the $5S$ ground state and to a scattering rate limited lifetime of the atomic ensemble. The maximum value of this light shift is achieved at the center of the beam and can be calculated to $\Delta \nu_{778}=(-97.6\times P)$ kHz/mW (repulsive potential) and $\Delta \nu_{1080}=(2.9\times P)$ MHz/W (attractive potential) for linearly polarized light \cite{grimm1999}. As the power of the fiber laser is at least three orders of magnitudes larger than the power of the diode laser, the light shift of the fiber laser exceeds that of the diode laser by more than a factor of 30 and thus dominates the light induced atomic dipole potential. The maximum scattering rate for atoms in the beam is calculated to be $\Gamma_{778}=(3.2\times P) s^{-1}$/mW and $\Gamma_{1080}=(0.47\times P) s^{-1}$/W. With typical beam powers of $P_{778}=300\mu$W and $P_{1080}=1$W the scattering rates become comparably large: $\Gamma_{778}=0.95 s^{-1}$ and $\Gamma_{1080}=0.47 s^{-1}$, limiting the atomic lifetime to about one second. Longer detection times can be obtained by either reducing the beam powers or by keeping the laser powers high but chopping the laser beams with a small duty cycle. The latter method ensures that the ionization still dominates the spontaneous decay during the detection periods, preserving the detection efficiency.

Once the ions are generated, they are captured by an ion optics and guided to a CEM-detector \cite{stibor2007}. As the ions hit the detector, secondary electrons are emitted causing an avalanche inside the CEM-tube. We observe the resulting charge pulse with standard amplification \cite{F100TD} and counting electronics \cite{SR430}.

\begin{figure}[tb]
\centerline{\scalebox{0.4}{\includegraphics{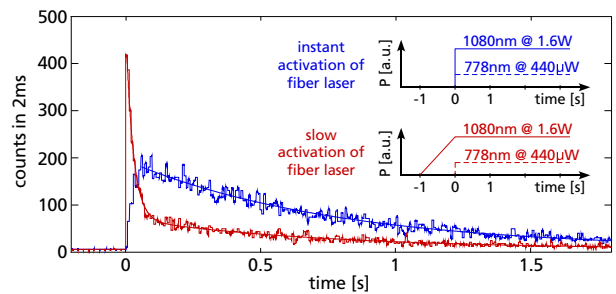}}}
\caption{(Color Online) Ion counting rate at the channeltron for different activation schemes of the lasers. If the fiber laser is activated synchronously with the diode laser, the counting rate decreases exponentially on a timescale of $\tau=771$ms (blue curve). Activating the fiber laser prior to the diode laser in a 1s linear ramp, the counting rate shows a double exponential decay (red curve). The atoms in the adiabatically loaded dipole trap are first ionized on a timescale of $\tau_1=27$ ms, whereas the remaining part of the cloud is ionized on a timescale of $\tau_2=724$ ms.}
\label{fig:Figure2}
\end{figure}

To demonstrate the operation of the detector, we prepare a cloud of about $7\times 10^5$ atoms at a temperature of $6 \mu$K underneath the hole in the chip. After instantaneously turning on the two ionization lasers with powers of $P_{778}=440\mu$W and $P_{1080}=1.6$W, we observe a counting rate at the CEM-detector which is shown by the blue curve in Fig.\;\ref{fig:Figure2}.
After a fast increase to its maximum value, the ionization rate decays exponentially on a timescale of $\tau=771$ms (blue curve fit). As the fiber laser forms an additional dimple trap inside the existing magnetic trap, we expect the activation time of the fiber laser to be crucial for the detected ionization rates. After linearly ramping up the power of the fiber laser within 1s, we observe a counting rate as shown by the red curve in Fig.\;\ref{fig:Figure2}. The slow activation of the dipole potential yields an adiabatic loading of atoms to the dipole trap which increases the density inside the ionization volume. Thus the initial ionization rate is much higher than in the non-adiabatic case. The atoms in the dipole trap are now ionized on a fast timescale of about $\tau_1=27$ms. Thereafter, only atoms from the magnetic trap passing the ionization volume are ionized, showing the same decay rate as in the non-adiabatic case $\tau_2=724$ms.

\begin{figure}[tb]
\centerline{\scalebox{0.4}{\includegraphics{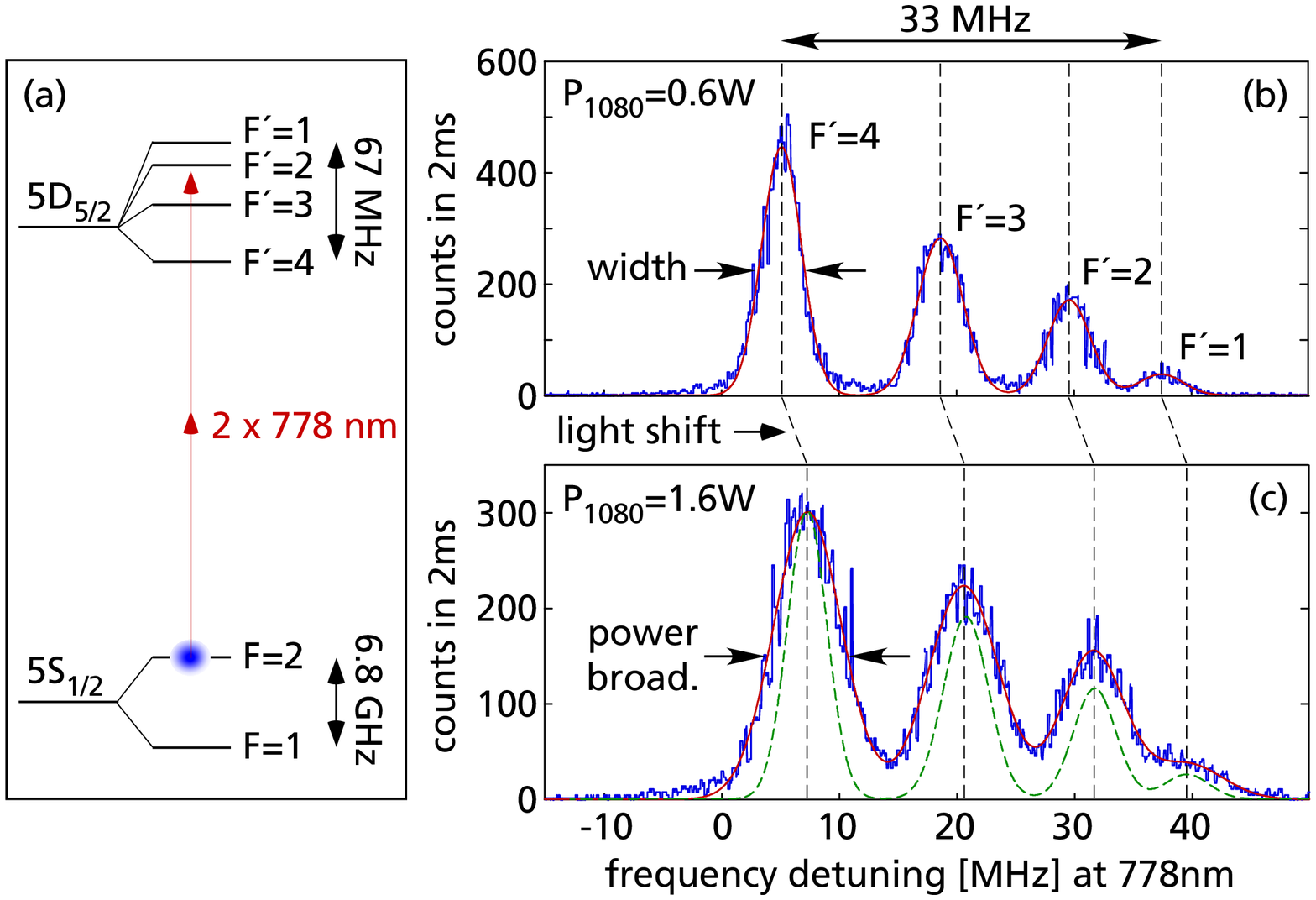}}}
\caption{(Color online) Scanning the two photon frequency over the hyperfine splitting of the $5D_{5/2}$-state (a) we observe resonance lines in the CEM signal, corresponding to an increased ionization rate for resonant two photon transitions [(b) and (c)]. The origin of the frequency detuning is chosen arbitrarily. The measured ion counting rate (blue curve) is described by a superposition of Gauss profiles, one for each resonance line (red curve). With higher fiber laser power, the spectral lines are light shifted to higher frequencies (see dashed lines) and power broadened. The power broadening is illustrated by the green dashed curve in (c), showing the spectrum of (b).}
\label{fig:Figure3}
\end{figure}

For optical spectroscopy we prepare a cloud of $1.8\times 10^6$ atoms at a temperature of $18\mu$K and instantaneously turn on the two lasers with the diode laser blue detuned to the two photon $F=2\rightarrow F'=1$ transition. As the atoms exhibit a strong attractive dipole potential from the fiber laser we wait 500ms for the atoms to thermalize into the newly generated potential configuration. For recording the spectra we continuously lower the frequency of the diode laser at a rate of $d\nu/dt=-45$MHz/s over a total range of 65MHz. As a frequency detuning $\Delta\nu$ of the diode laser corresponds to a detuning of $2\Delta\nu$ relative to the $5S-5D$ level spacing, we cover a total frequency interval of 130MHz including all hyperfine resonances of the two photon transition [Fig.\;\ref{fig:Figure3}(a)]. While scanning the diode laser frequency we monitor the counting rate at the CEM-detector. We repeat the experiment for 11 different powers of the fiber laser ranging from 80mW up to 1.6W, while keeping the diode laser power constant at $270\mu$W. Fig.\;\ref{fig:Figure3}(b) and (c) show the detected ion counting rate for two different powers of the fiber laser.
Whenever the two photon frequency is resonant to a $5D_{5/2}$ hyperfine transition, the ion production and thus the ion counting rate increases, showing clear resonance lines in the CEM signal. As the two photon transition allows $\Delta F=0,1,2$ we observe resonant transition lines for all four hyperfine levels of the $5D_{5/2}$-state. The spectra shown in Fig.\ref{fig:Figure3} are derived from a total of $5\times 10^4$ ion counts. With a detection efficiency of about $50\%$ \cite{stibor2007}, the spectra are thus taken with ionization losses of less than $7.5\%$ of the initial number of atoms.

For a quantitative analysis we model each resonance line with a Gaussian line profile. Fitting a superposition of four Gaussian profiles to the spectral data, we determine the spectral positions and widths of the four resonance lines for each power of the fiber laser. We found the frequency difference $\Delta\nu_{F',F'+1}$ between neighbored lines to be $\Delta\nu_{1,2}=(7.73\pm 0.15)$MHz, $\Delta\nu_{2,3}=(10.86\pm 0.11) $MHz, and $\Delta\nu_{3,4}=(13.49\pm 0.11) $MHz, which is in good agreement with literature values \cite{Nez1993}. Due to the AC-Stark effect the resonances are shifted linearly with the power of the fiber laser at a rate of $(2.4\pm 0.2)$MHz/W (at 778nm). Theoretically we expect a smaller rate of $1.5$MHz/W, which is probably due to an overestimation of the beam waist by 26\%. The spectral linewidth (FWHM) is also found to increase linearly with the fiber laser power by $(3.0\pm0.4)$MHz/W. We attribute this in parts to the increased ionization rate, lowering the effective 5D lifetime ($\sim 0.7$MHz/W) and to a power dependent line broadening due to an inhomogeneous light shift arising from the Gaussian beam profile of the fiber laser. The extrapolated zero power linewidth of $(2.6\pm0.4)$MHz is consistent with the different contributions from the natural transition linewidth ($\sim 0.33$MHz), the linewidth of the diode laser ($\sim 1.5$MHz) and the inhomogeneous Zeeman broadening due to the thermal spread of the atoms across the magnetic trap ($\sim 0.5$MHz).

\begin{figure}[tb]
\centerline{\scalebox{0.4}{\includegraphics{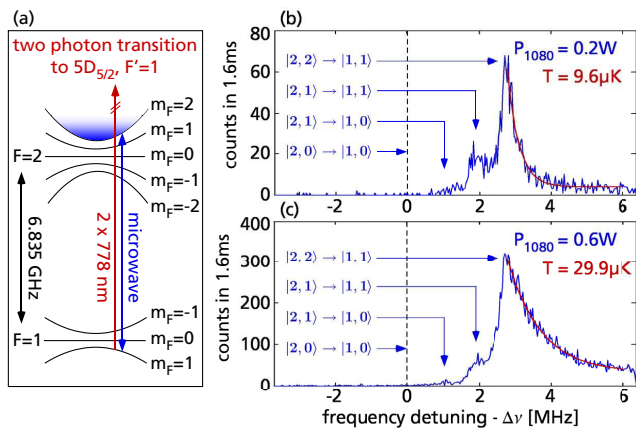}}}
\caption{(Color online) (a) The Zeeman-splitted $5S_{1/2}$ hyperfine states are coupled by a microwave near $6.8$GHz. Atoms in the $F=2, m_F=2$ state are coupled to the $F=1, m_F=1$-state from where they are ionized. (b)+(c) By scanning the microwave frequency we record the spectrum of the ion counting rate. It directly reveals the energy distribution of the atoms in the trap. The origin of the frequency axis is arbitrarily set to the $F=1,m_F=0 \rightarrow F=2,m_F=0$ transition. Additional structures indicated by the vertical arrows are due to residual population in the $F=2, m_F=1$ state.}
\label{fig:Figure4}
\end{figure}

For measuring the energy distribution of the atoms in the trap, we add a double resonance scheme to the existing setup [Fig.\;\ref{fig:Figure4}(a)]. The diode laser is tuned in resonance with the $5S_{1/2},F=1\;\rightarrow 5D_{5/2},F'=1$ transition. The atomic cloud is magnetically trapped in the $F=2, m_F=2$ groundstate and coupled to the $F=1,m_F=1$ groundstate with a tunable microwave near $6.8$GHz. Resonantly coupled atoms undergo a Rabi oscillation and get ionized by the lasers once they are in the $F=1$ state. Depending on the microwave frequency detuning $\Delta\omega$ (with respect to the resonance frequency for atoms at the magnetic trap bottom), the microwave addresses atoms at different Zeeman energies according to their position inside the magnetic field. For a parabolic magnetic trap this resonance shell is given by the surface of an ellipsoid defined by
\begin{equation}
\mu_B B(\bm{r})=\frac{1}{2}m\omega_x^2 x^2 + \frac{1}{2}m\omega_r^2 (y^2 + z^2)=\frac{2}{3}\hbar\Delta\omega
\end{equation}
with the Bohr magneton $\mu_B$ and the rubidium atomic mass $m$. For given $\Delta\omega$ all atoms on the resonance shell are coupled to the $F=1$ state, but only those from the intersection region with the laser beams are ionized. For very small frequencies ($\Delta\omega < 2\pi\times 1.3$kHz) the ellipsoid is fully covered by the laser beams. However, already for $\Delta\omega>2\pi\times 0.25$MHz, the resonance shell leaves the ionization volume in two dimensions, intersecting the laser basically in two single planar sheets at vertical positions of $z=\pm\sqrt{ \left( 4\hbar\Delta\omega/\left(3m\omega_r^2\right)\right)}$. With the atomic line density $n(z)=\int\!\!\int dx dy n(\bm{r})$ along the optical beam axis, the number of atoms in the resonance planes is given by:
\begin{equation}
n(z)dz\sim\frac{n\left(z\left(\Delta\omega\right)\right)}{\sqrt{\Delta\omega}}d\omega=n(\Delta\omega)d\omega.
\end{equation}
Scanning the microwave while monitoring the ion counting rate at the CEM thus allows for measuring the distribution function $n(\Delta\omega)$, which in case of a thermal cloud is given by
\begin{equation}
n(\Delta\omega)=\frac{\exp{\left(-\frac{2\hbar\Delta\omega}{3k_BT}\right)}}{\sqrt{\Delta\omega}}.
\label{eq:1}
\end{equation}
For large frequencies $\Delta\omega$ we thus expect an almost pure exponential behavior which allows for in situ temperature determination.

In contrast to the previous experiment we linearly ramp up the fiber laser within 500ms. After additional 500ms we turn on the diode laser ($P_{778}=300\mu$W) and the microwave (Rabi frequency of $2\pi\times 23.7$kHz) and tune its frequency from 6.841GHz to 6.835GHz within 320ms. Fig.\;\ref{fig:Figure4}(b) and (c) shows the monitored ion counting rate at the CEM detector for two different powers of the fiber laser.
As expected, starting from high detunings the counting rate increases exponentially, according to a Boltzmann factor with a temperature determined by an exponential fit [Eq.\ref{eq:1} and Fig.\;\ref{fig:Figure4}(b) and (c)]. At higher power, the dipole trap compresses the cloud more strongly and the temperature increases. This can be seen from Fig.\ref{fig:Figure4}(c) showing a larger fraction of atoms at higher microwave frequencies with respect to Fig. \ref{fig:Figure4}(b).

In summary we have demonstrated sensitive high resolution optical spectroscopy
of ultracold atoms trapped on a magnetic microchip. Decay curves and complete
optical spectra of a single cloud have been recorded within a few
hundred milliseconds and with little detection losses. The method offers a
promising new approach for studying atomic and molecular quantum gases.
Efficient recording of three body decay rates, for instance, is
crucial for the investigation of recently discovered Efimov physics
\cite{kraemer2006}. Spectroscopy of trapped cold atoms with single atom
sensitivity would also help for observing many body quantum states with non
classical number fluctuations \cite{Jo2007} and for developing quantum sensors
and quantum computation on atom chips \cite{trupke2007}. Furthermore, high
resolution spectroscopy near surfaces \cite{Failache1999,fichet2007,nayak2007} can now be
extended to quantum gases. Finally, optical spectroscopy of small samples of
trapped atoms is discussed in the context of testing fundamental physics with
antihydrogen \cite{gabrielse2008}. One and two photon spectroscopy of trapped
atoms has already been developed for large samples of atomic hydrogen
\cite{cesar1996,eikema2001}.

We thank S. Kraft for early contributions to the experiment and acknowledge support by the Deutsche Forschungsgemeinschaft and the Landesstiftung Baden-W\"urttemberg.

\bibliographystyle{apsrev}

\end{document}